\documentclass{article}
\usepackage{epsfig}
\newcommand{\bfr}{\begin{flushright}}
\newcommand{\efr}{\end{flushright}}

\newcommand{\ifb}{\mbox{$\rm fb^{-1}$}}
\newcommand{\kt}{\mbox{$k_{\rm t}$}}
\newcommand{\akt}{\mbox{$k_t$}}
\newcommand{\pt}{\mbox{$p_{\rm T}$}}
\newcommand {\et} {\mbox{$E_{\rm T}$}}
\newcommand {\Ht} {\mbox{$H_{\rm T}$}}
\newcommand {\epem} {\mbox{$\rm e^+e^-$}} 
\newcommand {\mpmm} {\mbox{$\rm \mu^+\mu^-$}}

\begin{document}
\title{Production of electroweak bosons in association with jets with the ATLAS detector 
\thanks{Presented at  the Low x workshop, June 13-18 2017, Bari, Italy}%
}
\author{S. Spagnolo\footnote{stefania.spagnolo@le.infn.it} ~on behalf of the ATLAS Collaboration
\\
{\small
INFN Lecce and 
Dip. di Matematica e Fisica ``Ennio De Giorgi", Universit\`a del Salento
}
\smallskip\\
}
\date{October 23rd, 2017
}
\maketitle
\begin{abstract}
This report summarises studies of associated production of electroweak gauge bosons and jets in proton-proton collisions at LHC with center of mass energy of 8 and 13 TeV. It is based on a selection of results published by the ATLAS Collaboration in the first half of 2017. 
\\
~
\\
PACS number(s): 
\end{abstract}

\section{Introduction}
This report collects results of analyses performed mainly on 20.2~\ifb\ of $pp$ collisions at $\sqrt s=8$~TeV collected in 2012 by the ATLAS detector \cite{detATLAS}. A few more recent studies use the ATLAS data set recorded in 2015 when LHC has been running at the $pp$ center of mass energy of $13$~TeV. 

In particular, a first set of results refers to studies of isolated photons with jets at 8 TeV, of $Z$ bosons with jets at 13 TeV, and of the splitting scale of the anti-\akt\ jet clustering algorithm in events with a $Z$ boson at center of mass energy of 8 TeV.  
All these studies exploit the electroweak (EW) probes to test Quantum Chromo Dynamics (QCD) in a very clean environment; most of them are updates at high center of mass energy of classical QCD measurements at LHC, that provide great input for the validation of calculations and the tuning of Monte Carlo~(MC) generators. 

A second set of results comprises a study at 8~TeV of the angular separation of a $W$ boson from the closest jet, targeting real $W$ emission, and measurements of the production cross Section of $W$+2jets and $Z$+2jets in the vector boson fusion (VBF) kinematic region, the first at 8~TeV and the latter at 13~TeV.  In this case, the focus is on measurements and modeling of EW physics, however the results provide very interesting insight on QCD in unconventional phase space regions.

$W$ and $Z$ are reconstructed in the leptonic decay channels with $e^\pm$ and $\mu^\pm$, photons are reconstructed from electromagnetic clusters without a matching track, or from $e^+e^-$ pairs with a common vertex inside the detector from conversions. To allow an easy comparison with theory predictions, all these analyses are producing cross sections at particle level, corrected for detector resolution and acceptance, in  a fiducial region tailored to the detector acceptance.

\section{QCD measurements}
\subsection{Photons + jets}
\label{photons+jets}
This process proceeds at leading order (LO) through two different mechanisms: direct production of a photon in the hard scattering ($q g \rightarrow q \gamma$) and fragmentation of a coloured parton to a high transverse energy, \et, photon. Although the separation is theoretically meaningful only at LO, the different underlying dynamics can be probed with some specific variables, like the scattering angle in the center of mass frame ($\theta^\star$), whose distribution depends on the spin of the propagator in the t-channel scattering. In addition, the distribution of the QCD radiation around the photon and the leading jet is sensitive to the different color flow between initial and final state particles in the two production modes. 
Central isolated photons of  \et\ greater than 130 GeV are used along with jets of $p_{\rm T} > \rm 130,~65~and~50~GeV$ for the leading, sub-leading and next to sub-leading jet to measure differential distributions of several variables: photon and jet \pt, invariant mass of the photon-jet system, scattering angle, separation in azimuth between photon and jets and between jets, and $\beta$ angle, that will be defined later. 
State of the art MC generators with LO matrix elements (ME) interfaced to parton shower (PS) and QCD next to leading order (NLO) calculations are compared to data. 

The differential inclusive photon plus at least one jet cross section is measured with high resolution over five orders of magnitude as a function of \pt\ of the leading jet and transverse energy of the photon. The NLO predictions from {\tt Jetphox}, which accounts both for the direct and the fragmentation production modes, are in very good agreement with data over the entire phase space explored. The LO MC generators {\tt Sherpa} and {\tt Pythia} are also describing well the data overall, with the exception of the high \et\ region, where they both overestimate the data. The distribution of $\cos \theta^\star$ is measured in nine intervals of the invariant mass of the photon and the leading jet. When comparing data with LO predictions obtained separately with {\tt Jetphox} for the direct production and the fragmentation production modes, it is clearly apparent that neither of the two is able to describe the data properly.  On the other hand, an excellent agreement between the complete {\tt Jetphox} NLO QCD prediction and data is observed, implying that both a good description of the dynamics and a satisfactory modelling of its scale dependence are achieved. 
\begin{figure}[htb]
\centerline{%
\begin{tabular}{cc}
\includegraphics[width=5cm]{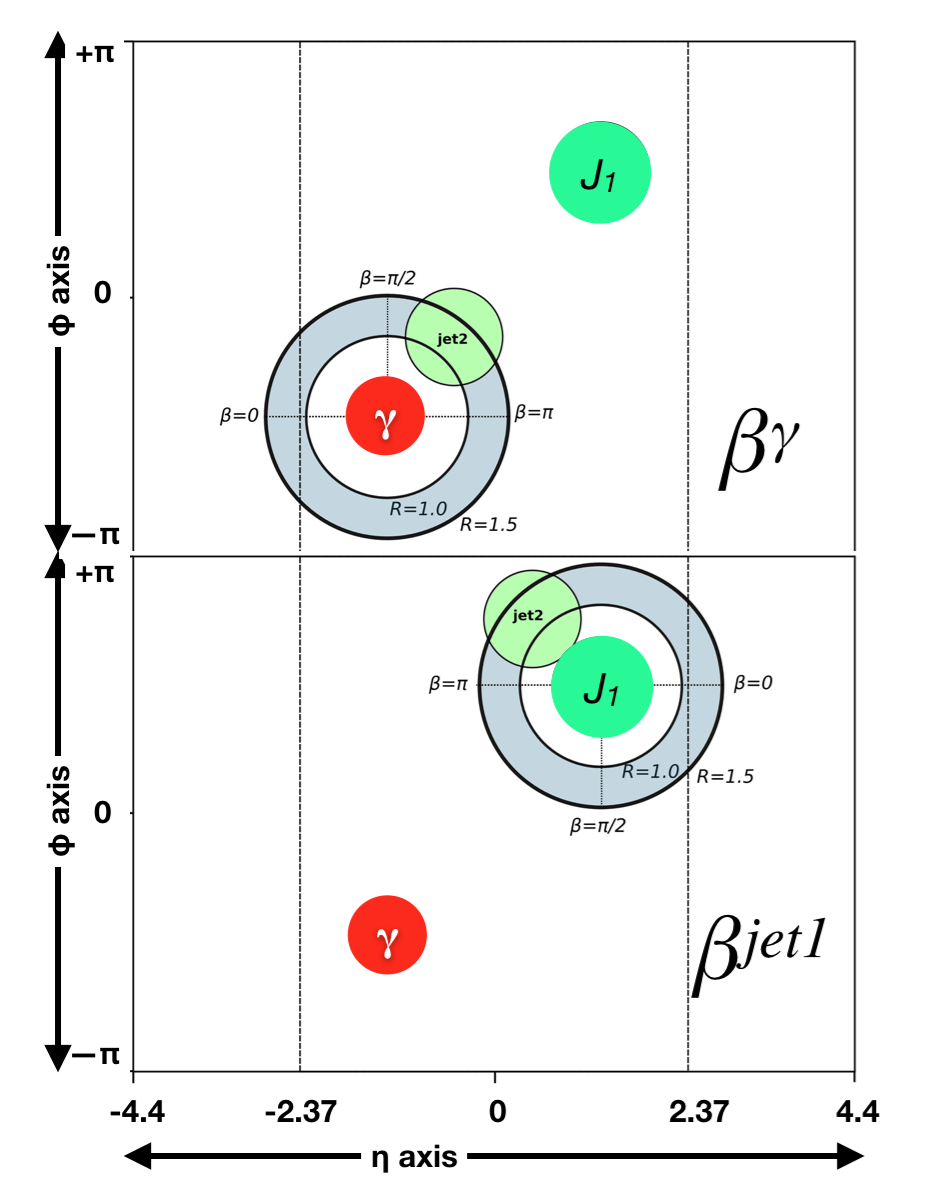} & 
\includegraphics[width=6.cm]{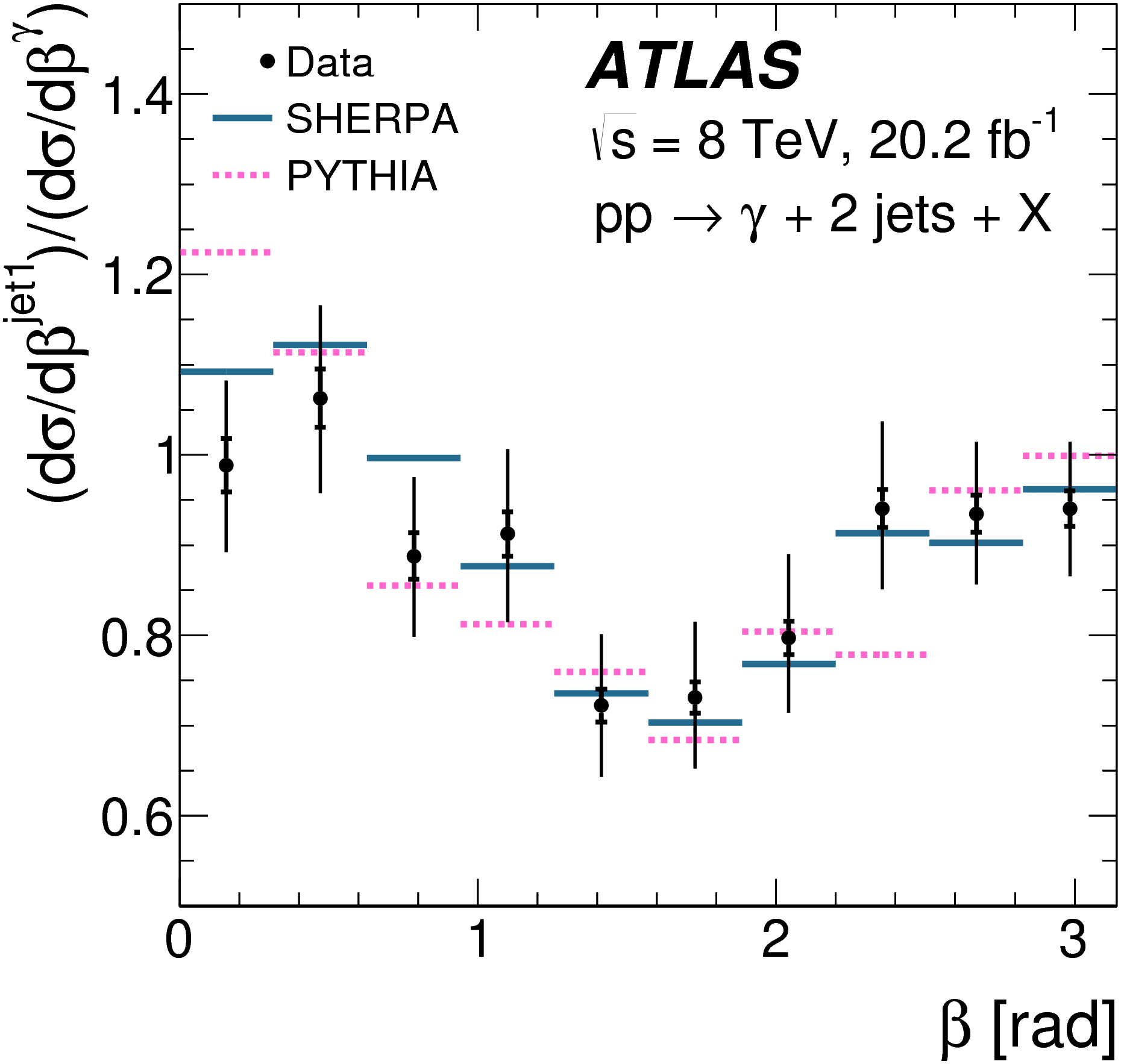} \\
\end{tabular}
}
\caption{On the left a sketch illustrating the geometrical meaning of the angular variable $\beta$, as defined with respect to the photon ($\beta^{\gamma}$, upper plot) and to the leading jet ($\beta^{jet1}$, bottom plot). The measurement of $\beta^{\gamma}$ ($\beta^{jet1}$) is performed only for sub-leading jets lying in the annular region in grey around the photon, $\gamma$ (the leading jet, $J_1$).  On the right the ratio of the differential cross sections as a function of $\beta^{jet1}$ and $\beta^{\gamma}$, measured in Ref.~\cite{citeGammaj}, is compared with predictions from MC generators.
\label{figGammaj}}
\end{figure}

The measurements of the differential production cross section of a photon and at least 2 or at least 3 jets show the limitations of the LO MC generators. In the case of these observables, NLO QCD predictions, obtained with {\tt BlackHat+Sherpa}, describe well the data with some tension at high transverse energy of the photon. {\tt Pythia8} is poorly describing most of the distributions, while {\tt Sherpa}, using LO ME for up to 4 jets, performs quite well with the exception of the domain of high transverse energy of the photon. 

An interesting and peculiar measurement presented in Ref.~\cite{citeGammaj} is the differential cross section as a function of a newly defined variable, $\beta$, meant to measure the $\eta-\phi$ proximity of the sub-leading jet ($jet2$) to the photon or to the leading jet ($jet1$).  The sketch on the left of Figure \ref{figGammaj} helps visualising the definition: 
$$
\beta^x = tan^{-1} \frac{|\phi^{jet2}-\phi^{x}|}{sign(\eta^x)\cdot (\eta^{jet2}-\eta^x)}
$$
where $x$ can be $\gamma$ or $jet1$.
The $\beta$ variable is measured only in events with at least two jets and with a value of $\Delta R=\sqrt{\Delta\phi^2+\Delta\eta^2}$ (the distance between the sub-leading jet and the photon, for the case of $\beta^{\gamma}$, and between the sub-leading jet and the leading jet, for $\beta^{jet1}$) in the interval between 1 and 1.5. This selection requirement preserves photon isolation and prevents any overlap of the samples of events used for the differential cross section measurements as a function of $\beta^\gamma$ and of $\beta^{jet1}$. The ratio of the two cross sections is shown in the plot on the right of Figure \ref{figGammaj}. It clearly exhibits a non-flat trend, implying a different pattern of radiation around the photon and the leading jet. The MC generators provide a good description of this behaviour, that is clearly expected but it is measured unambiguously for the first time in this ATLAS analysis. 

\subsection{$Z$ + jets at 13~TeV}
\label{Zjets13}
A study of $Z$+up to 7 jets has been performed with the data at $\sqrt{s}=13$~TeV collected in 2015 \cite{citeZjets}. Similar results were published in 2012 at 7~TeV for $Z$+jets \cite{citeZjets7} and  for $W$+jets~\cite{citeWjets7}.  
\begin{table}[htb]
\centerline{%
\includegraphics[width=12cm]{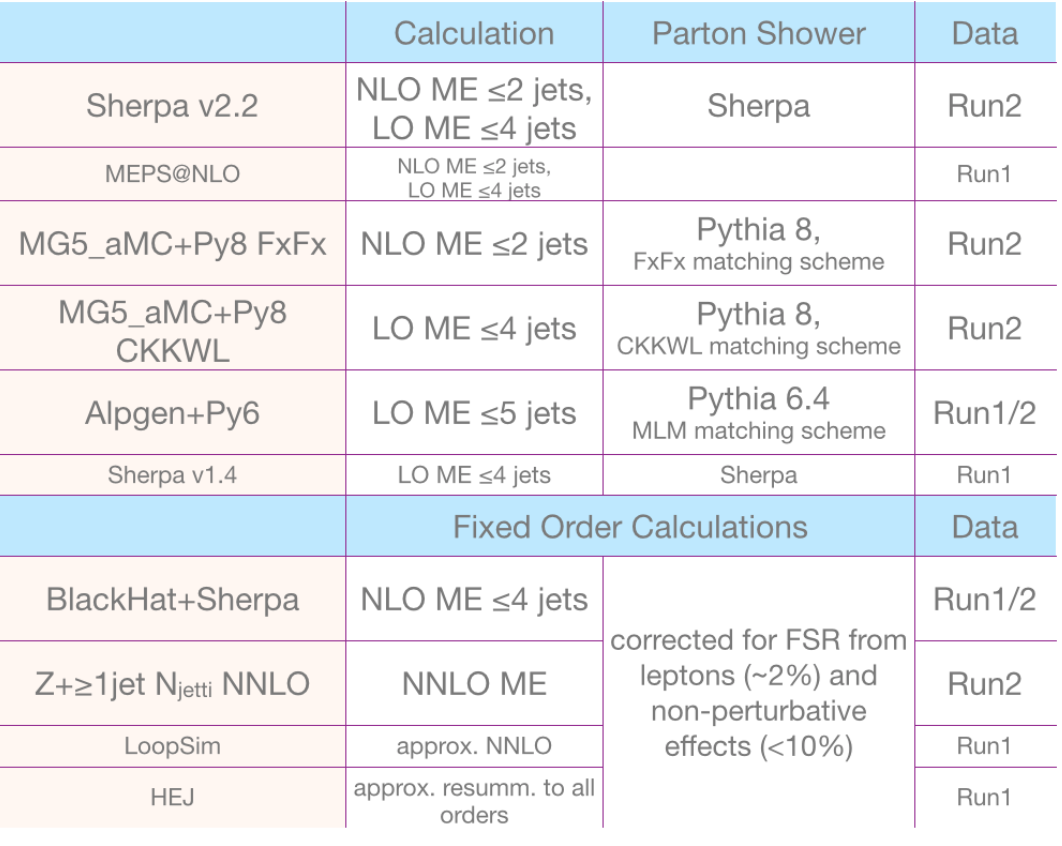} 
}
\caption{Main features of the generators and calculations predicting the production of gauge bosons in association with jets. The measurements presented in Ref.~\cite{citeZjets} are compared to the predictions  listed here as used for the {Run}2 data. \label{tabPred}}
\end{table}
They are clean measurements for accurate testing of QCD calculations and a precious playground for tuning of the MC generators which are extensively used to model accurately these important backgrounds for searches and measurements of rare SM processes. The steady progress in the theory tools for LHC is fully exploited in the recent paper, where a large set of precise MC generators and fixed order calculations, whose main features are summarised in Table \ref{tabPred}, is compared with the ATLAS measurements. 
\begin{figure}[htb]
\centerline{%
\begin{tabular}{cc}
\includegraphics[width=6cm]{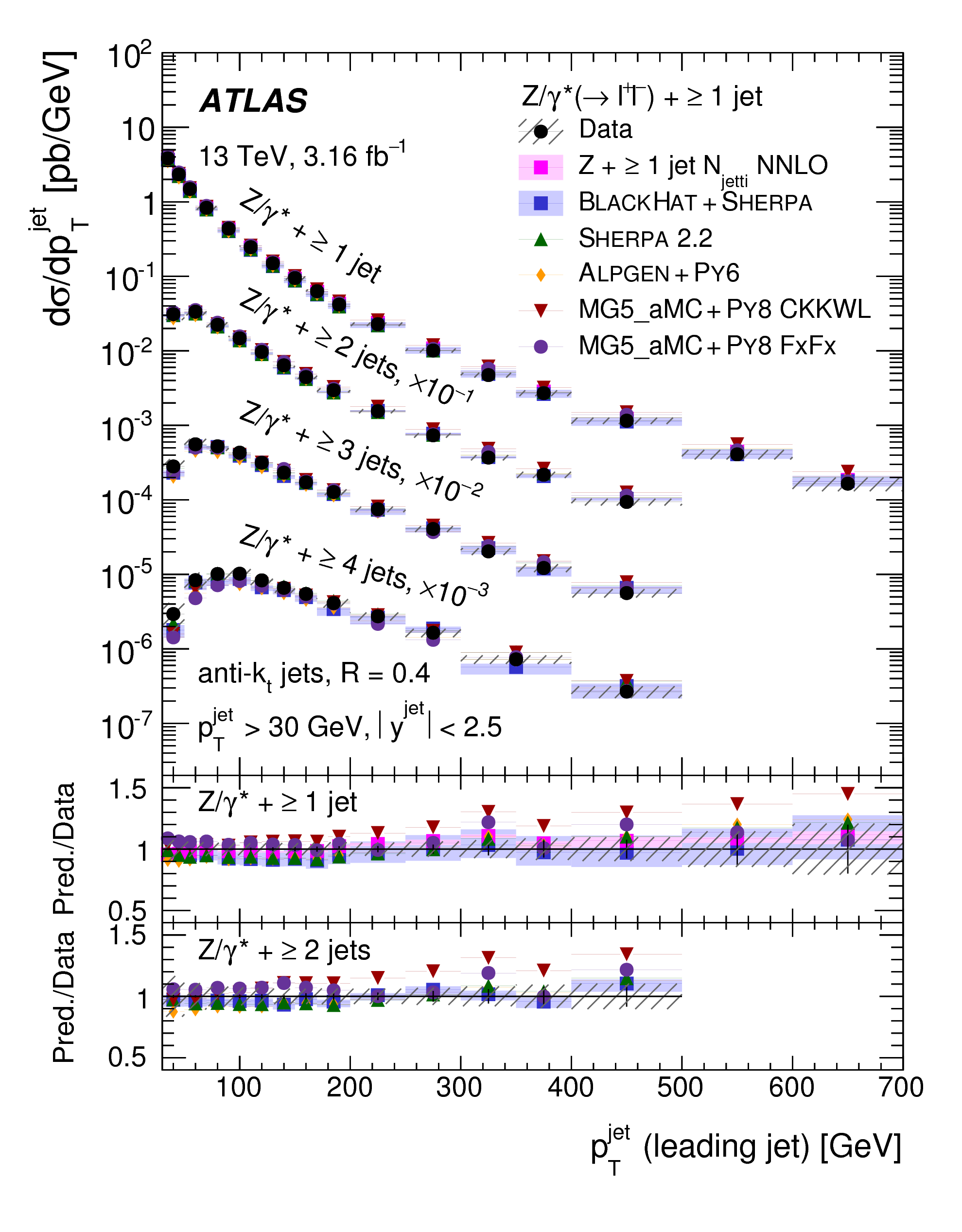} & 
\includegraphics[width=6cm]{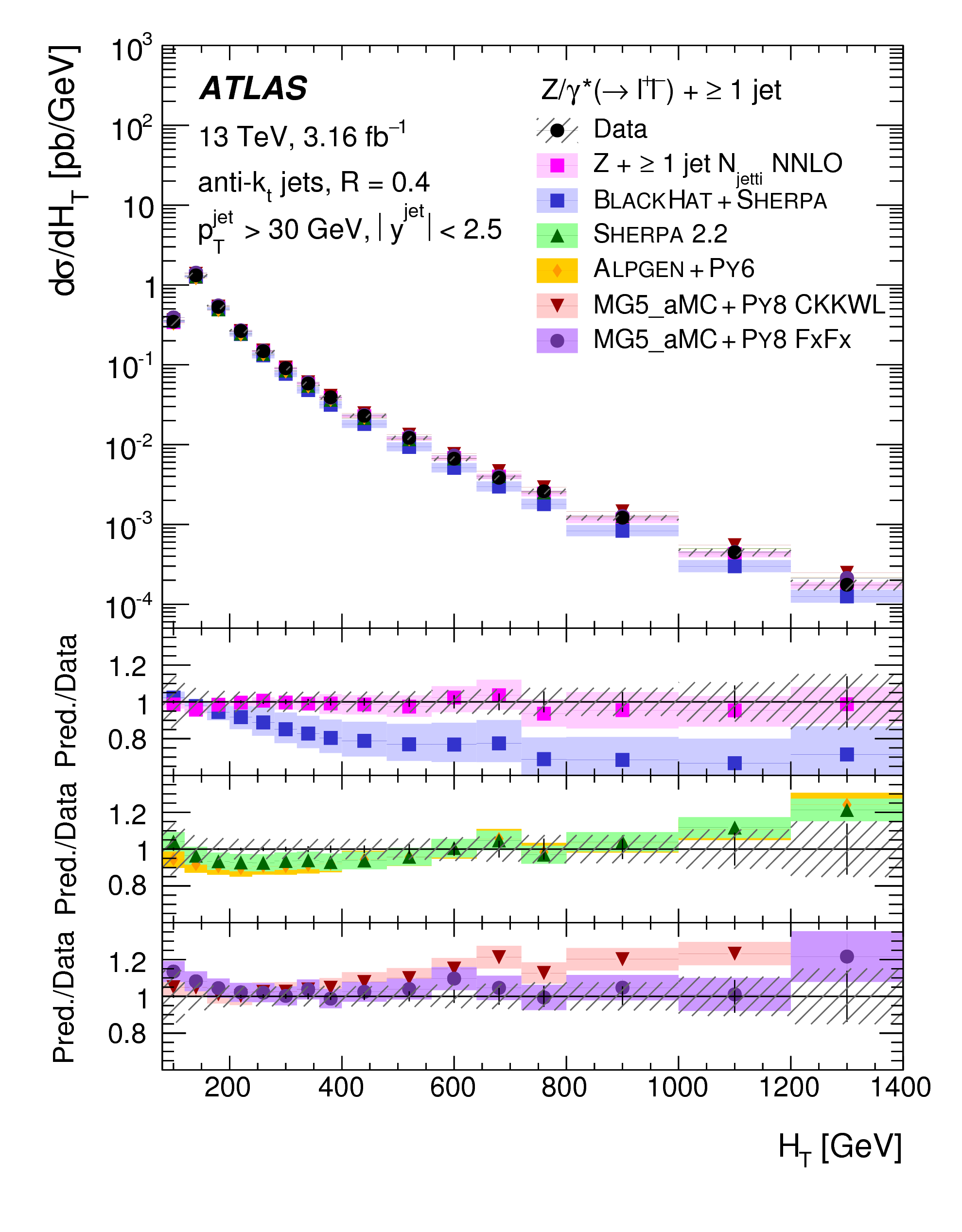} \\
\end{tabular}
}
\caption{Differential cross section as a function of leading jet \pt\ for various jet multiplicities in $Z$+jets events (left). Distribution of the \Ht\ variable (right) of  $Z$+jets events. Both plots are extracted from Ref.~\cite{citeZjets}. \label{figZjets}}
\end{figure}

The observables investigated are inclusive and exclusive jet multiplicities, \pt\ and rapidity of the leading and sub-leading jets, \Ht, angular variables and di-jet invariant mass. 

The differential cross section as a function of jet multiplicity is compared to four MC generators and one fixed order calculation. In the high jet-multiplicity regime some systematic discrepancy between data and all MC generators is observed, although the size of the offset is of the order of the total experimental error.  

The jet \pt\ is a key kinematical variable in all LHC analyses. In Figure \ref{figZjets}, on the left, the differential cross section as a function of the leading jet \pt\ is shown for many jet multiplicities. All predictions based on NLO ME perform well; the same is true for the LO generator {\tt Alpgen}. The LO {\tt Madgraph} prediction instead overestimates the cross section above 200 GeV, confirming an already observed problem, presumably related to the dynamic scale choice implemented in this generator. 
The next to next to leading order (NNLO) calculation, based on the $N$-jettiness subtraction scheme, represented by the pink dots in Figure \ref{figZjets}, predicts very well the inclusive cross section for $Z$+jets. 
On the right of Figure \ref{figZjets} another crucial variable is shown: the scalar sum of the transverse momentum of jets and leptons, \Ht. The value of \Ht\ is commonly used as renormalisation and factorisation scale. Moreover, \Ht\ is often used as a discriminating variable in searches for new phenomena leading to a reach final state. Here the fixed order calculation needs the NNLO accuracy to match data. MC generators interfaced with PS are not really accurate over the entire spectrum. 
Some other distributions measured in Ref.~\cite{citeZjets} will be presented later for comparison with the results in the selection of $W$ + at least 2 jets produced via electroweak processes. 

\subsection{\kt\ splitting scale at 8~TeV}
\label{ktsplit}
An interesting and recent study of the \kt\ splitting scale in events with jets and a $Z$ boson is reported in Ref.~\cite{citeKtSplit}.  It provides insight into QCD with a different approach with respect to the standard gauge boson plus jets measurements and it gathers important evidence for the need of more tuning of QCD MC. 
The jet resolution scale of the anti-\akt\ algorithm has been measured at different orders (from 0 to 7) in events with a $Z$ reconstructed in the \epem\ or \mpmm\  decay using all charged tracks from the primary vertex with \pt $> 400$~MeV and pseudo-rapidity below 2.4. Low scales probe hadronization, fragmentation, underlying event modeling and parton shower, while high scales probe perturbative QCD. Differential cross section as a function of $\sqrt d_i$ (for $i=0, ..., 7$) are measured with two choices of the radius parameter $\Delta R=0.4$ and $\Delta R=1$. 
\begin{figure}[htb]
\centerline{%
\begin{tabular}{cc}
\includegraphics[width=6cm]{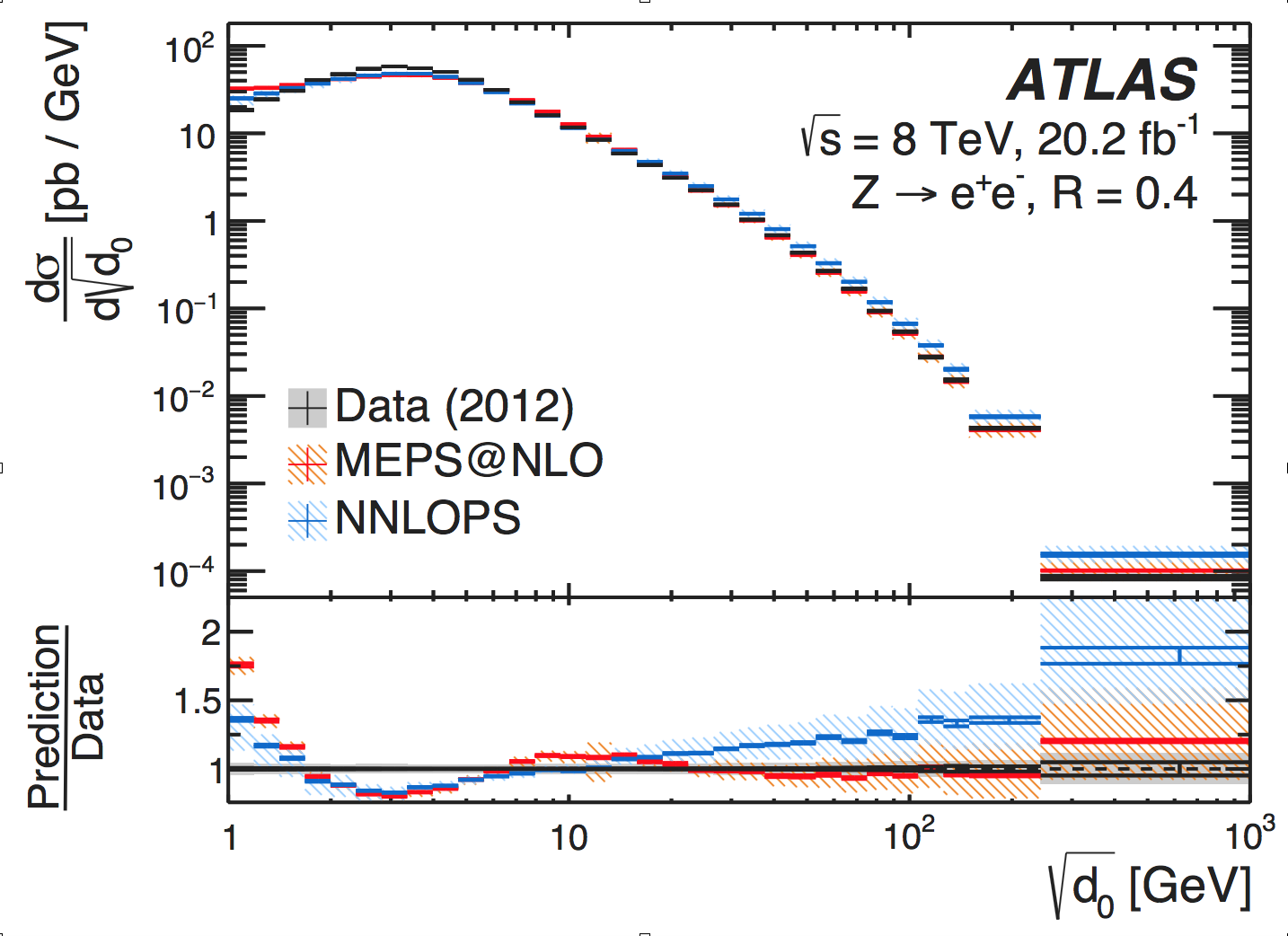} &
\includegraphics[width=6cm]{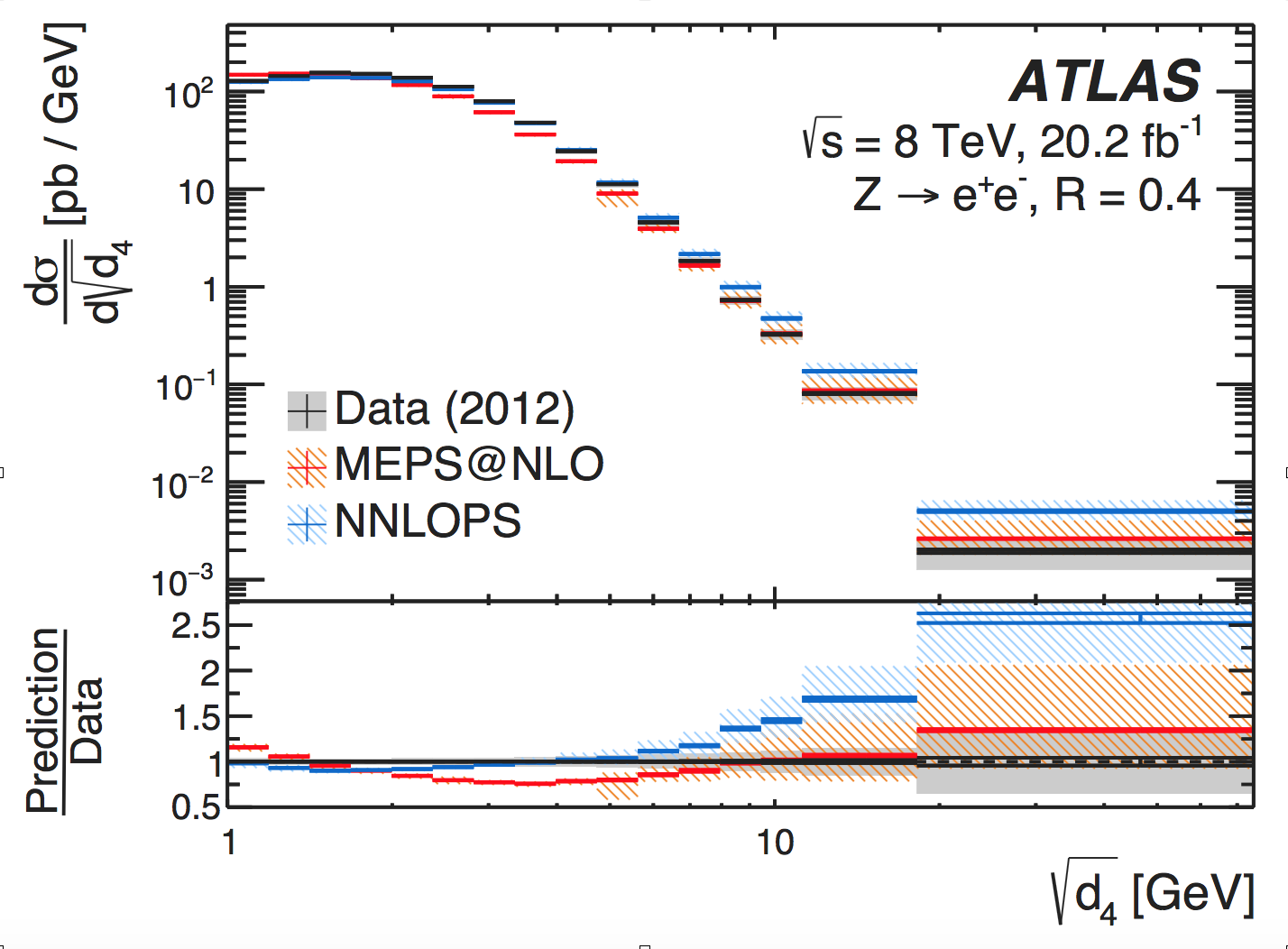} \\
\end{tabular}
}
\caption{Differential cross section as a function of splitting scale at order 0 (left) and at order 4 (right) for the anti-\akt\ jet clustering algorithm in events with $Z$ and jets. The plots are a selection of the measurements presented in Ref.~\cite{citeKtSplit} compared with two state of the art theory predictions. \label{figKtSplit}}
\end{figure}

In Figure \ref{figKtSplit} there are some examples of cross sections. Measurements are compared to state of the art MC generators: {\tt MEPS@NLO}, a prediction implemented in {\tt Sherpa} using a NLO ME for up to 2 jets and LO matrix elements for up to 4 jets, and  {\tt NNLOPS}, a NNLO calculation matched to the PS of {\tt Pythia8}. 
Surprisingly none of the generators provides a satisfactory description at all orders and all scales.  In particular, at low-order ($d_0$, $d_1$) both predictions overshoot data at low scales and fall 20\% below data in the region around 3~GeV. At higher orders, $d_{k>2}$,  {\tt NNLOPS} is in good agreement with data in the regime of soft hadronic activity. In the perturbative region at all orders {\tt NNLOPS} overestimates data while  {\tt MEPS@NLO} is in good agreement with the measurements.  

\section{QCD and EW measurements}
\subsection{Angular distribution of $W$ events with high \pt\  jets at  8~TeV}
\label{collWprod}
An analysis of the angular separation between a $W$ boson and a high \pt\ jet is presented in Ref.~\cite{citeCollW} with the aim of studying the radiation of a real $W$ off a quark in a pure QCD event. Therefore, the final state of interest consists  of two high \pt\ jets (back to back to first approximation) and a $W$ close to one of them, reconstructed through the muon decay.  The topology of these events is similar to that of t-tbar events in the high boost regime. 
\begin{figure}[thb]
\centerline{%
\includegraphics[width=9cm]{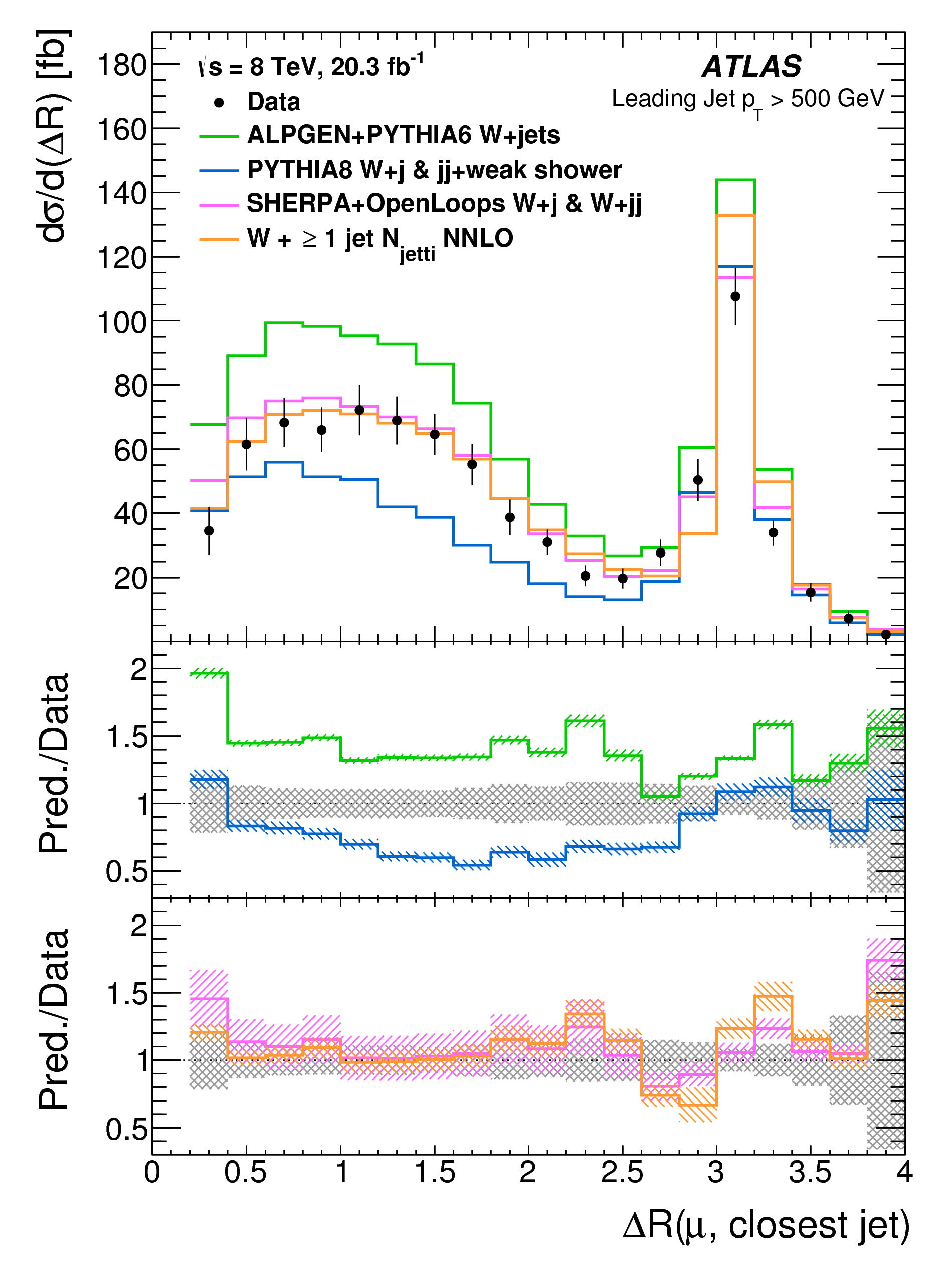} 
}
\caption{Differential $W$+jets cross section as a function of the separation $\Delta R$ between the $\mu$ from the $W$ decay and the closest jet  for the selection of events discussed in Ref.~\cite{citeCollW}.  \label{figCollW}}
\end{figure}
The collinear divergence in the cross section of a radiative process, which holds for a massless particle, translates into an enhancement of the cross section for high transverse momentum of the jets (when $m_{W}$ can be considered small compared to the momentum of the quarks). This study allows to test the modelling of this process, which can be described in various ways, including a formalism similar to the Sudakos Parton Showering. 
Several MC generators and fixed order predictions are compared to data. The process competes with the standard $W$+jet production where, at first order, the W is produced back to back with a jet and a second jet is radiated from an initial or final state parton. 
Therefore the selection is based on the request of a distance in $\eta-\phi$ between an isolated muon (from the $W$ decay) and the closest jet (of \pt\ $> 100$ GeV) smaller than 2.4 and greater than 0.2 (corresponding to the size of the isolation cone around the muon) and the presence of a jet with \pt\ $> 500$ GeV.

The measured cross section integrated in the collinear $W$ region (i.e. $0.2 < \Delta R < 2.4$) is reported in Table \ref{tabCollW}. 
Figure \ref{figCollW} shows the differential cross section as a function of $\Delta R$ between the muon and the closest jet obtained from the distributions subtracted of the expected background and corrected for detector effects. The normalisation of the main backgrounds (top events, multi jet events with a muon from a jet and $Z$+jets with the $Z$ decaying to \mpmm)  are adjusted using data control regions where a purity, in the targeted background, above 90\% is achieved  by reversing the selection criteria used in the nominal selection to suppress it. In the distribution, one can clearly distinguish the region of collinear $W$ emission from the region of direct production of the $W$ in the hard scattering. The comparison with theory shows that {\tt Alpgen}, using LO ME for up to 5 jets, is overestimating the cross section, although this is true also in the pure QCD $W$+jet region. The {\tt Pythia8} prediction, combining a LO ME for $W$+jets and a LO ME for $jj$ with weak showering,  is not good enough to describe the data.  The fixed order predictions based on ME which are at least NLO in QCD are equally good.  Another interesting observation is the enhancement of the cross section at increasing \pt\ of the jets in the collinear region, a remnant of the collinear divergence. 
\begin{table}[tb]
\centerline{%
\includegraphics[width=12cm]{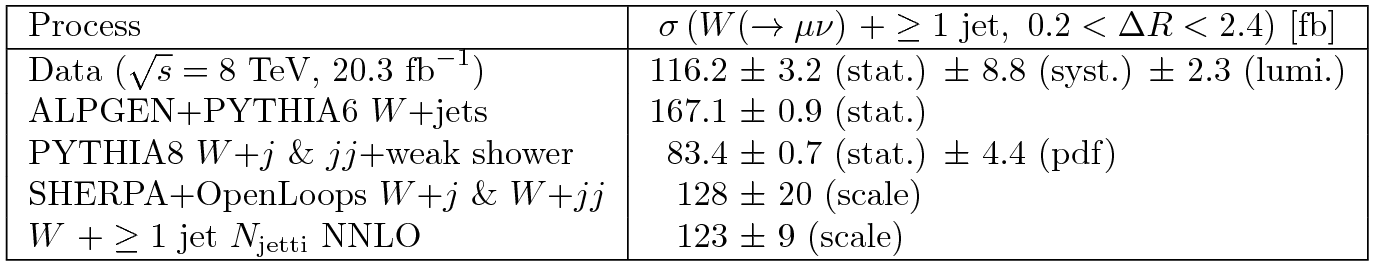} 
}
\caption{Fiducial cross section for the collinear emission of a $W$ boson measured in Ref.~\cite{citeCollW} and compared to several theory predictions. \label{tabCollW}}
\end{table}

\subsection{Electroweak production of $Wjj$ at 8 TeV }
\label{eqWjjprod}
Electroweak production of $Wjj$ at 7 and 8 TeV is studied in Ref.~\cite{citeEWWjj} and  limits are derived on anomalous trilinear gauge boson couplings. The analysis is meant to isolate the pure electroweak (EW) contributions to $W$ + 2 jets production. It follows a similar measurement of EW production of $Z$ + 2 jets at 8 TeV published in 2014 \cite{citeEWZjj8}. 
The main focus is clearly on the VBF contribution, the top-left diagram in Figure \ref{figdigramsWjj}, which probes the Gauge structure of the SM.  However, to isolate the EW contribution, described by the three top diagrams in Figure  \ref{figdigramsWjj}, the selection strategy must fight against a 10 times higher cross section from processes that are due to both EW and QCD interactions (some examples are shown in the bottom diagrams in Figure  \ref{figdigramsWjj}) and that can even interfere with the EW VBF scattering amplitude. Therefore, this measurement critically depends on the accurate modelling of the QCD contribution. 
The analysis is based on a preliminary inclusive selection of a VBF-like final state with a lepton from the $W$, well separated from two high \pt\ jets with a large rapidity gap $\Delta y_{12}$ between them and a large di-jet invariant mass.
\begin{figure}[thb]
\centerline{%
\begin{tabular}{ccc}
\includegraphics[width=3.0cm]{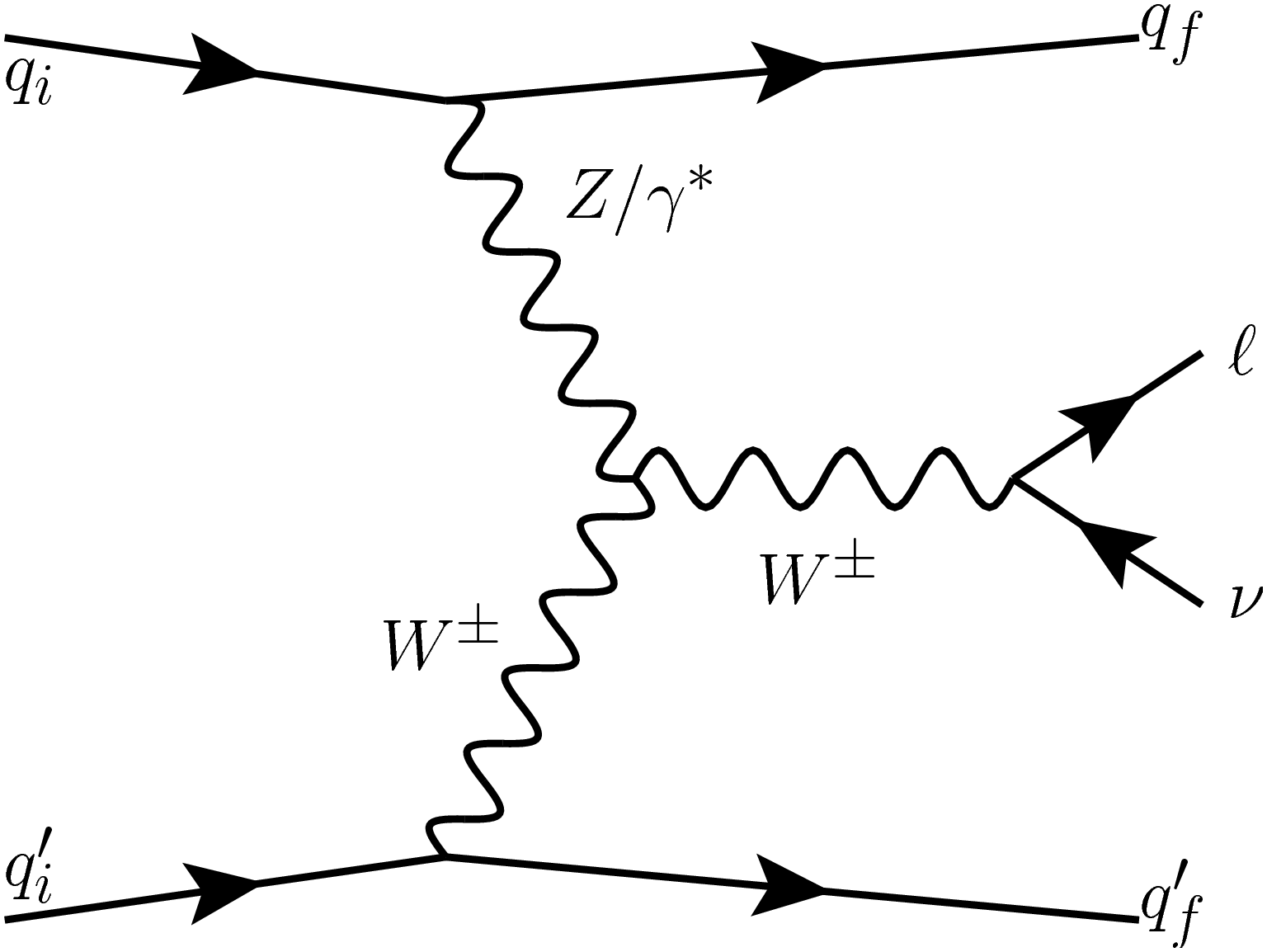} &  
\includegraphics[width=3.0cm]{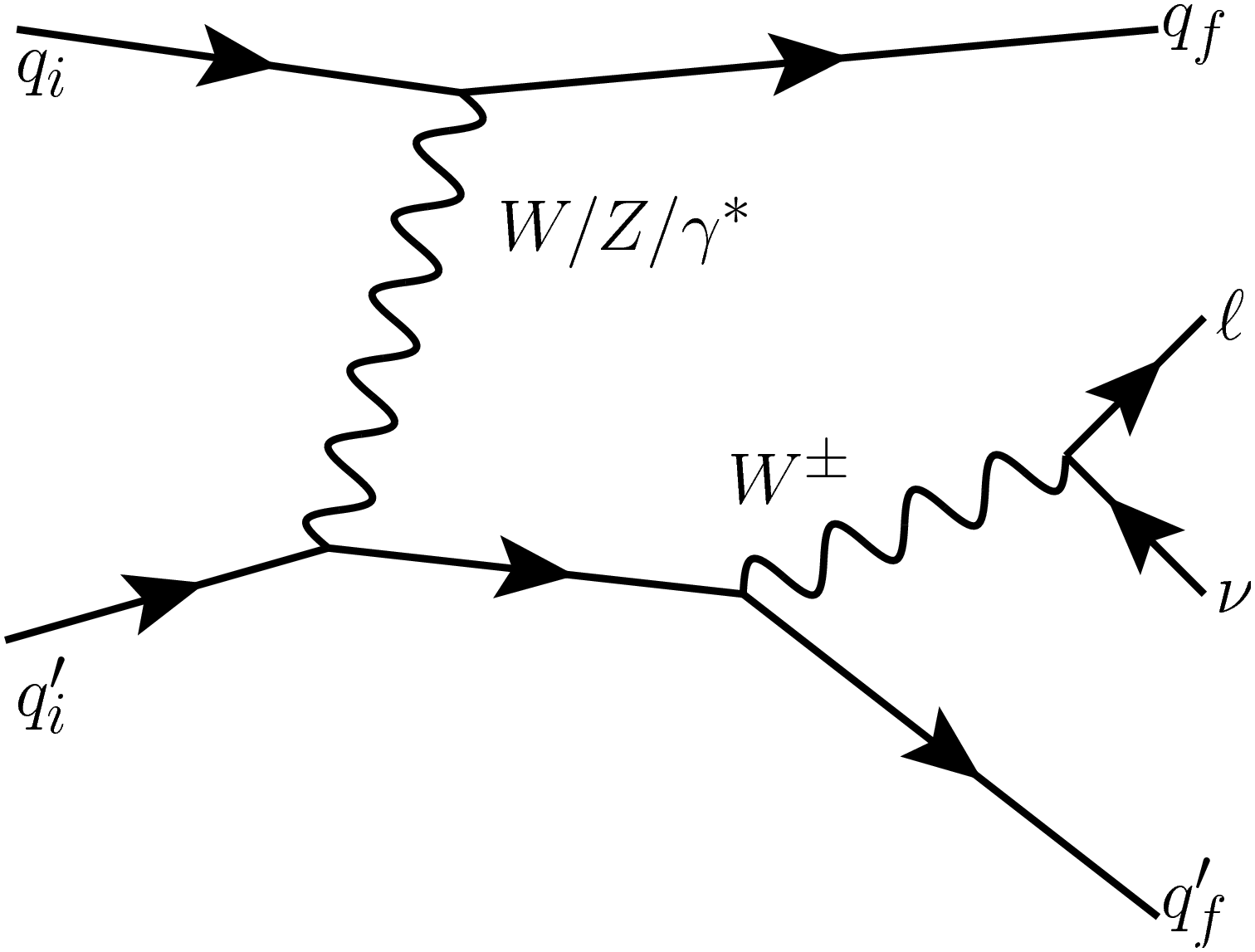} &  
\includegraphics[width=3.0cm]{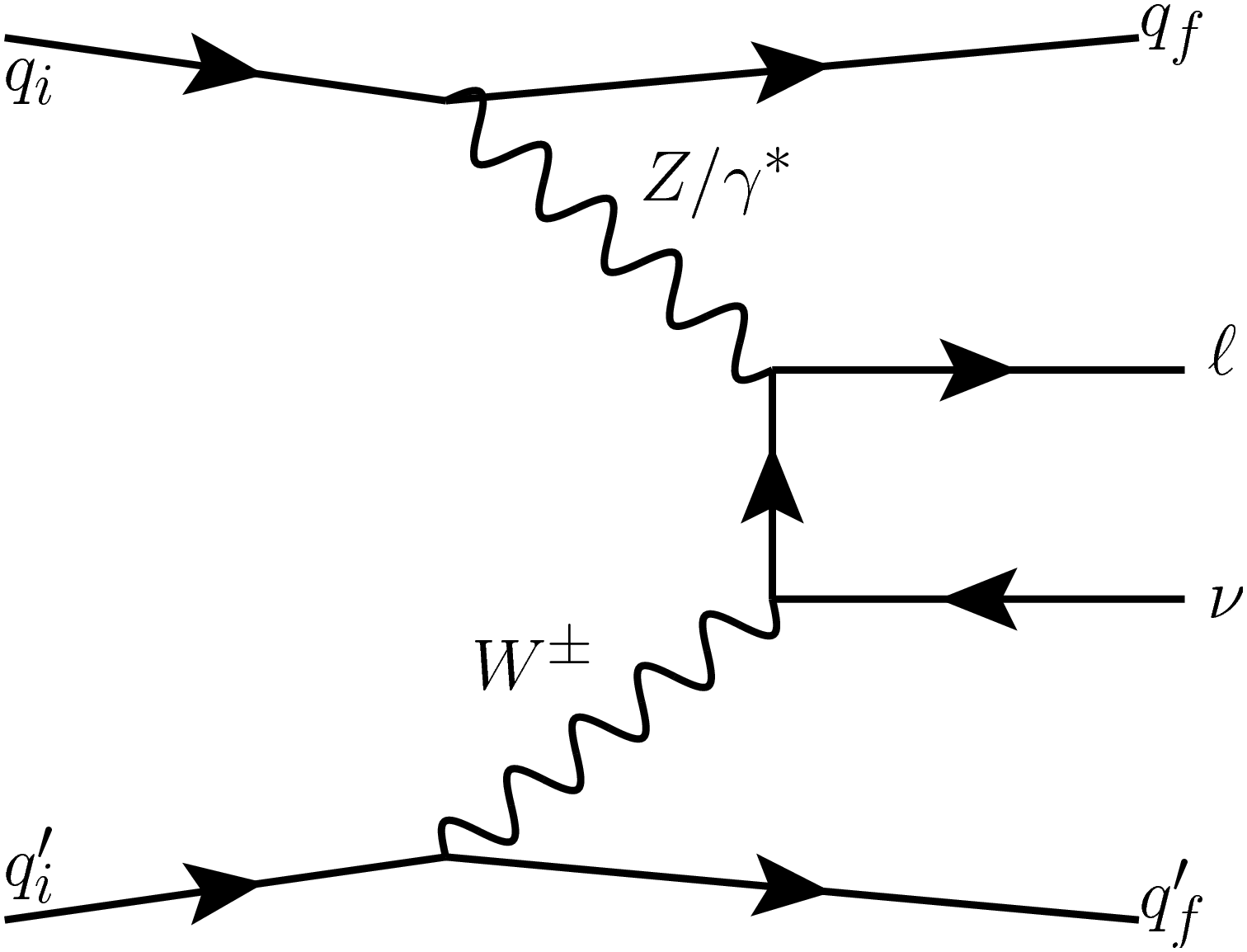} \\
\includegraphics[width=3.0cm]{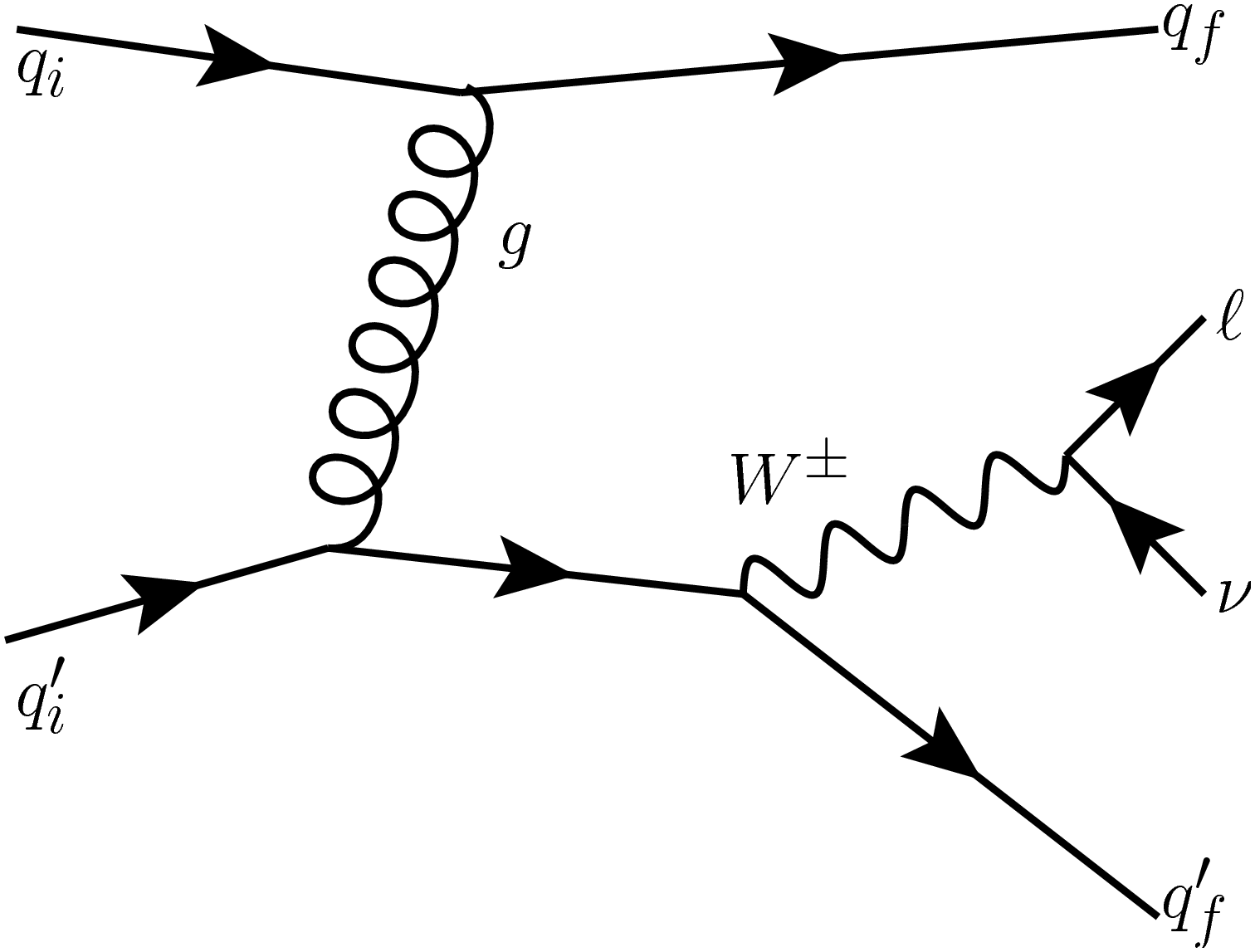} &  
\includegraphics[width=3.0cm]{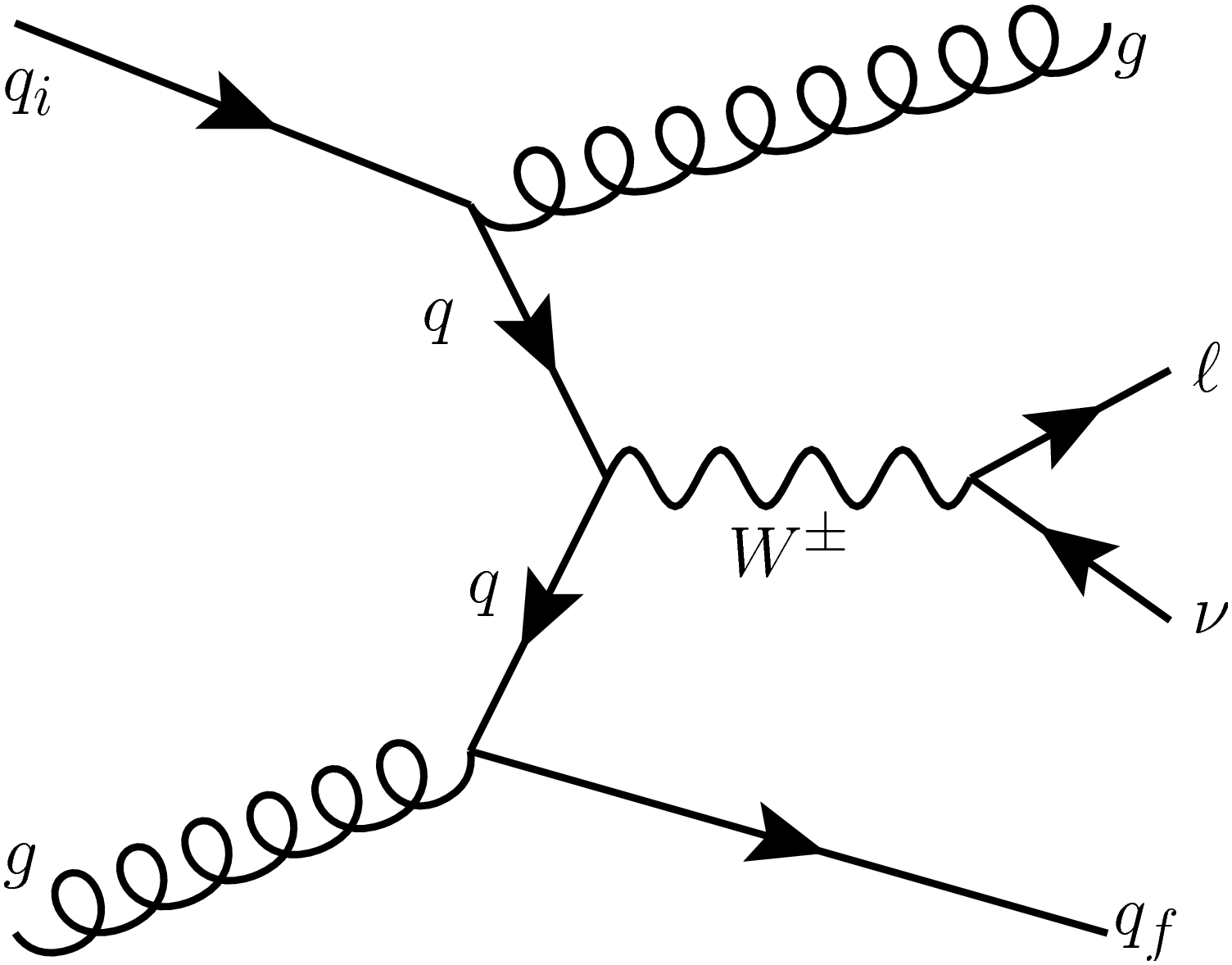} &  
\includegraphics[width=4.5cm]{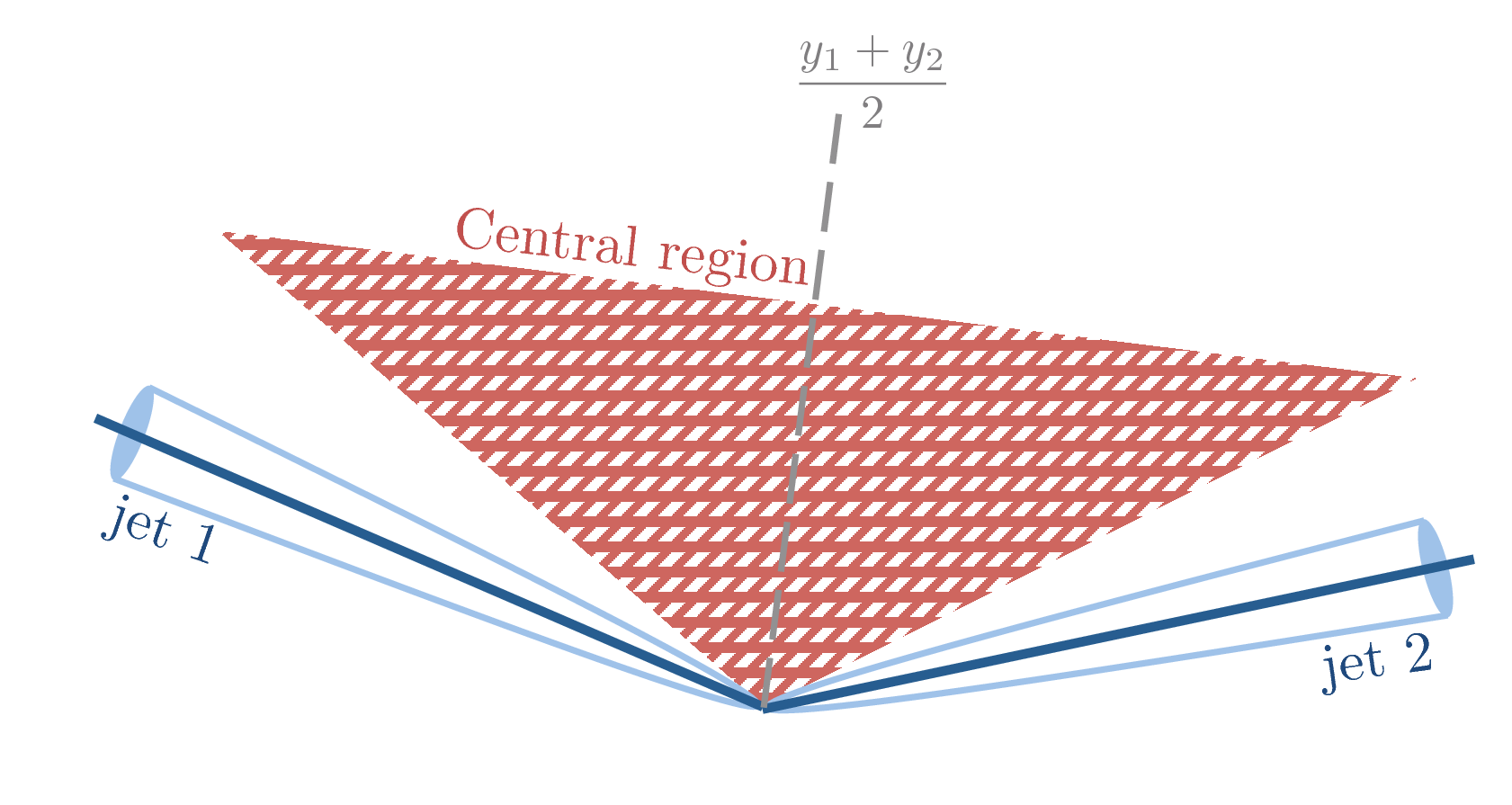} \\ 
\end{tabular}
}
\caption{The three Feynman diagrams at the top represent the production of $W$+2 jets through pure electroweak interactions; at the bottom the two diagrams show examples of QCD production of the same final state. In the bottom right corner the sketch illustrates the definition of the ``centrality" variable which is used in the analysis reported in Ref.~\cite{citeEWWjj} to define kinematic regions with enhanced, or suppressed, electroweak production of the $Wjj$ final state. \label{figdigramsWjj}}
\end{figure}
Then the events are classified into four categories based on the centrality of the lepton and of any extra jet in the event. The centrality is defined as the absolute value of the difference between the lepton (or  jet) rapidity and the average rapidity of the two leading jets, normalised to the di-jet rapidity gap $\Delta y_{12}$. The diagram in the bottom-right corner of Figure \ref{figdigramsWjj} shows as a shaded red area the region corresponding to ``central'' objects. The EW contribution to the inclusive event selection defined by the general cuts is concentrated at values of the lepton centrality below 0.4. Therefore, the EW enhanced region is defined by the presence of a central lepton and no central jets. On the other hand, events without both central jets and central leptons or with both a central jet and a central lepton are very rich in QCD $Wjj$ events and are used as control region and validation region, respectively, to constrain with data the QCD contribution to $Wjj$ in the EW enhanced signal region. In particular, the QCD $Wjj$ control region is used to determine the shape of the di-jet invariant mass spectrum for this category of events, which is later summed, with a floating normalisation, to the corresponding EW component for the fit to the distribution measured in the EW enhanced region. This treatment of the data allows to control the systematics from data modeling and to obtain the most precise measurement so far at LHC of a pure EW cross section based on the data set collected at 8 TeV.
Alongside with this measurement and with the resulting limits on anomalous trilinear gauge couplings, a number of detailed differential cross section measurements are produced in the inclusive as well as in the QCD enriched regions which are a great testbed for QCD predictions in the extreme phase space configuration explored by VBF and vector boson scattering processes. 
\begin{figure}[thb]
\centerline{%
\begin{tabular}{cc}
\includegraphics[width=6.0cm]{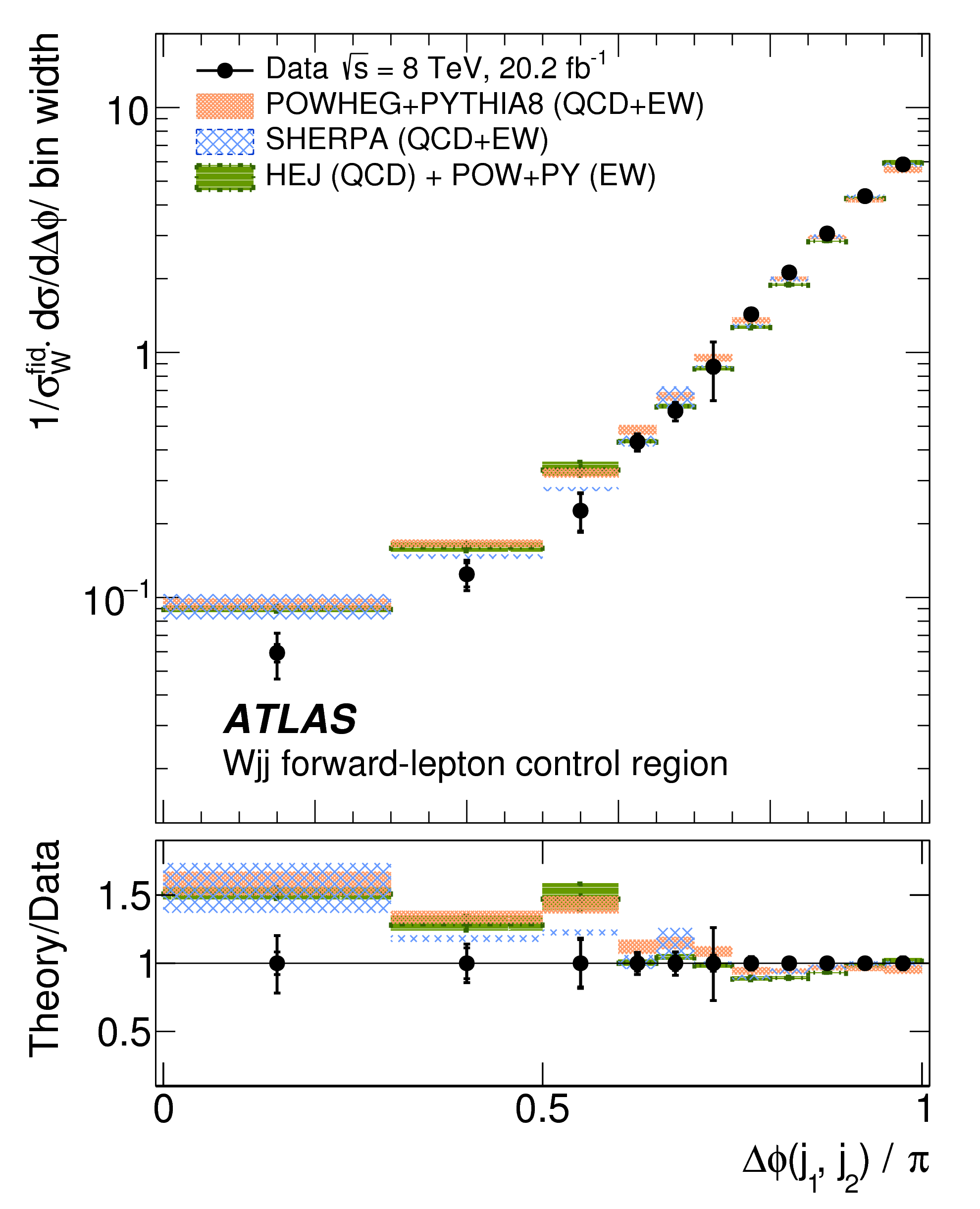} &  
\includegraphics[width=5.8cm]{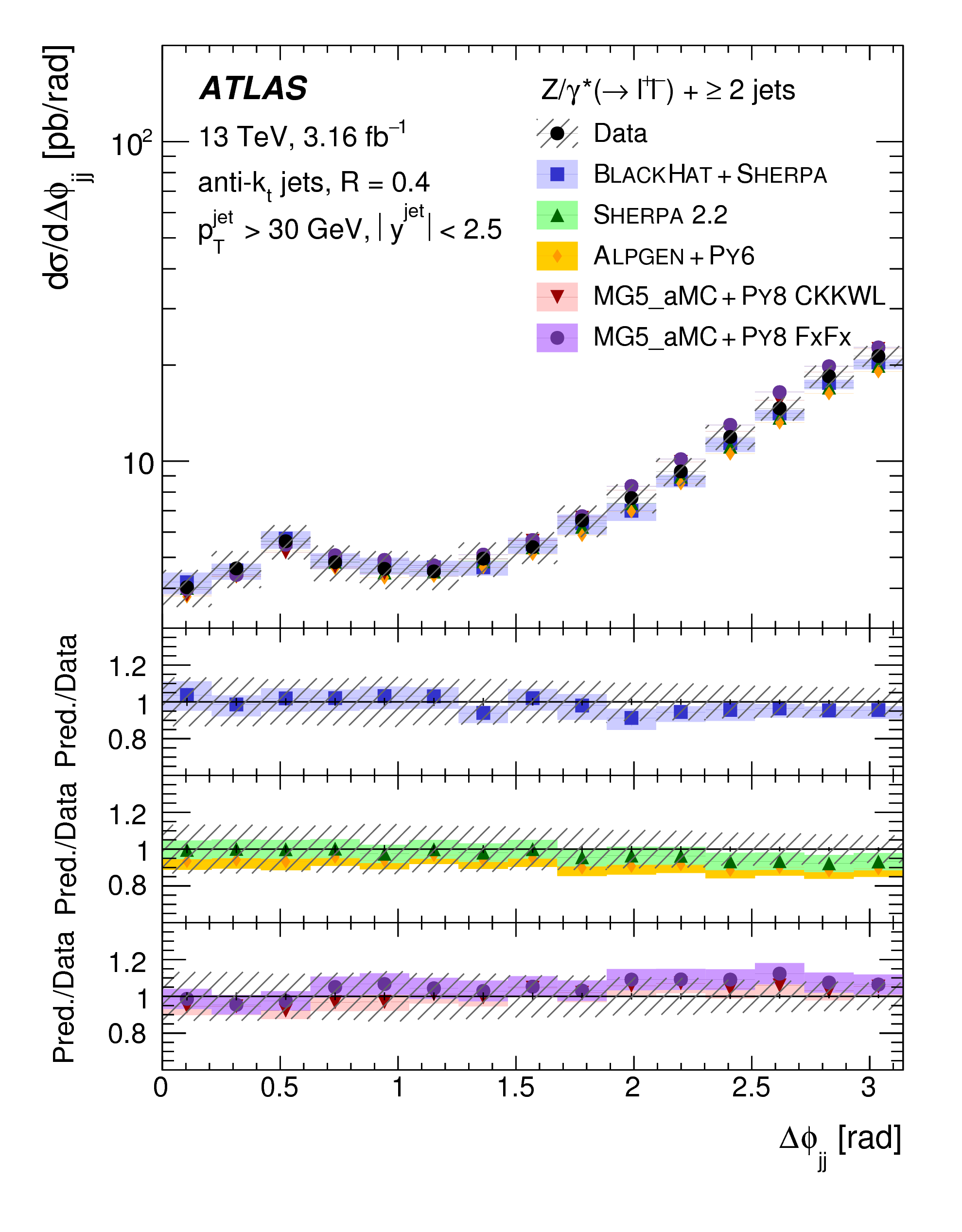} \\
\end{tabular}
}
\caption{On the left, the normalised differential cross section as a function of the azimuthal separation between the two leading jets in events of the QCD control region passing the inclusive selection for electroweak $Wjj$ production described in Ref.~\cite{citeEWWjj}. On the right, the differential cross section as a function of the azimuthal separation between the two leading jets in a standard selection of a gauge boson ($Z$) and two jets described in Ref.~\cite{citeZjets}. \label{figDPhi2pap}}
\end{figure}

A selection of these measurements is shown here and compared to the corresponding measurements performed in the standard $Z$ + jets analysis at 13~TeV, described earlier in Section \ref{Zjets13}, for the categories of events with a $Z$ boson and at least two jets. 
In Figure \ref{figDPhi2pap},  the distance in the azimuthal angle between the leading jets, a quantity that allows to test the balance of soft and hard emission in the generators, is shown. On the left the distribution measured in the $Wjj$ selection in a VBF phase space and in a region enriched in QCD contributions is displayed. The predictions are all overshooting data at low $\Delta \phi$, while in the standard phase space explored by the $Z$ + 2 jets analysis all MC generators and fixed order predictions are very good in describing the data, as can be seen in the right plot of Figure \ref{figDPhi2pap}. 
\begin{figure}[thb]
\centerline{%
\begin{tabular}{cc}
\includegraphics[width=6.0cm]{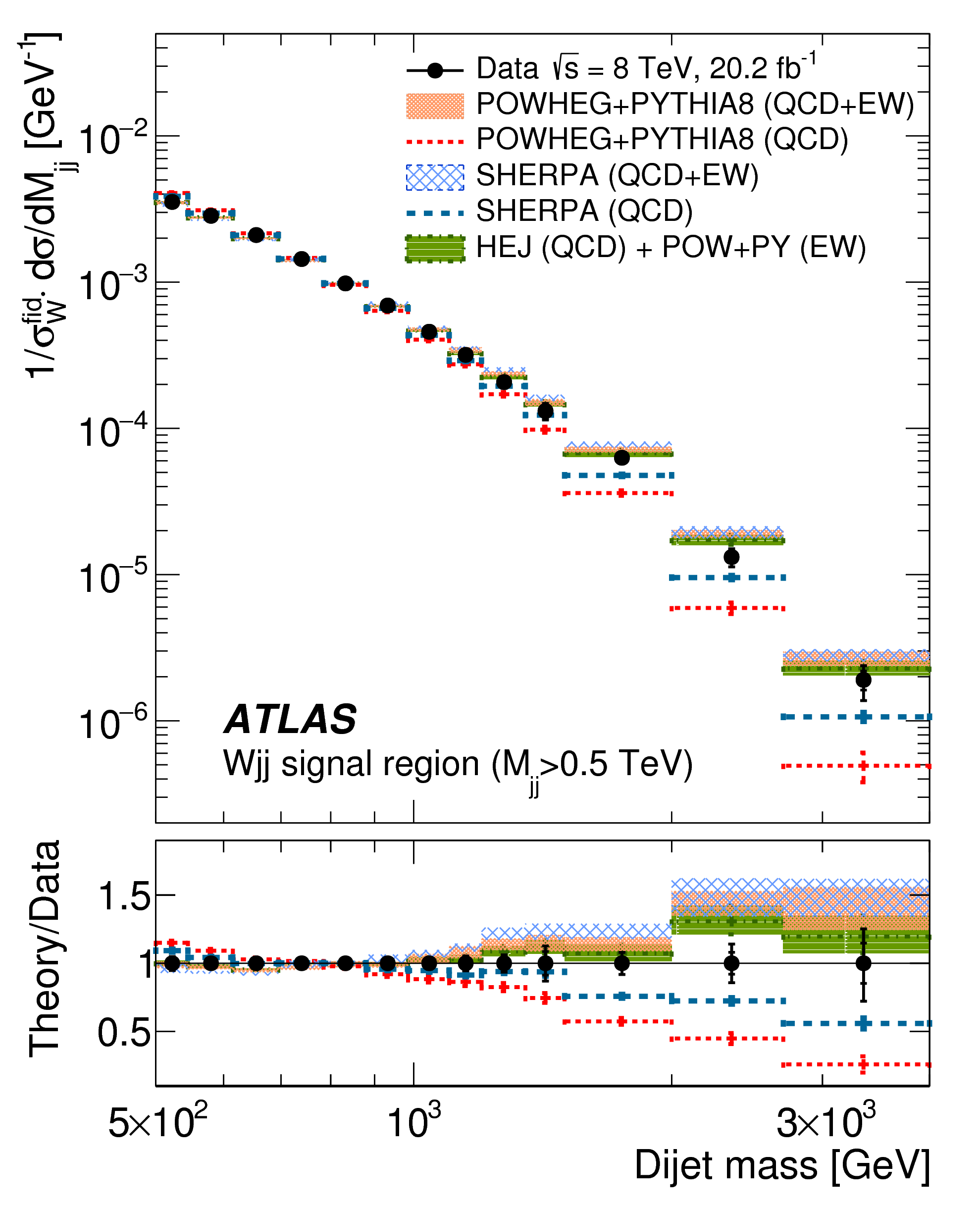} &  
\includegraphics[width=5.8cm]{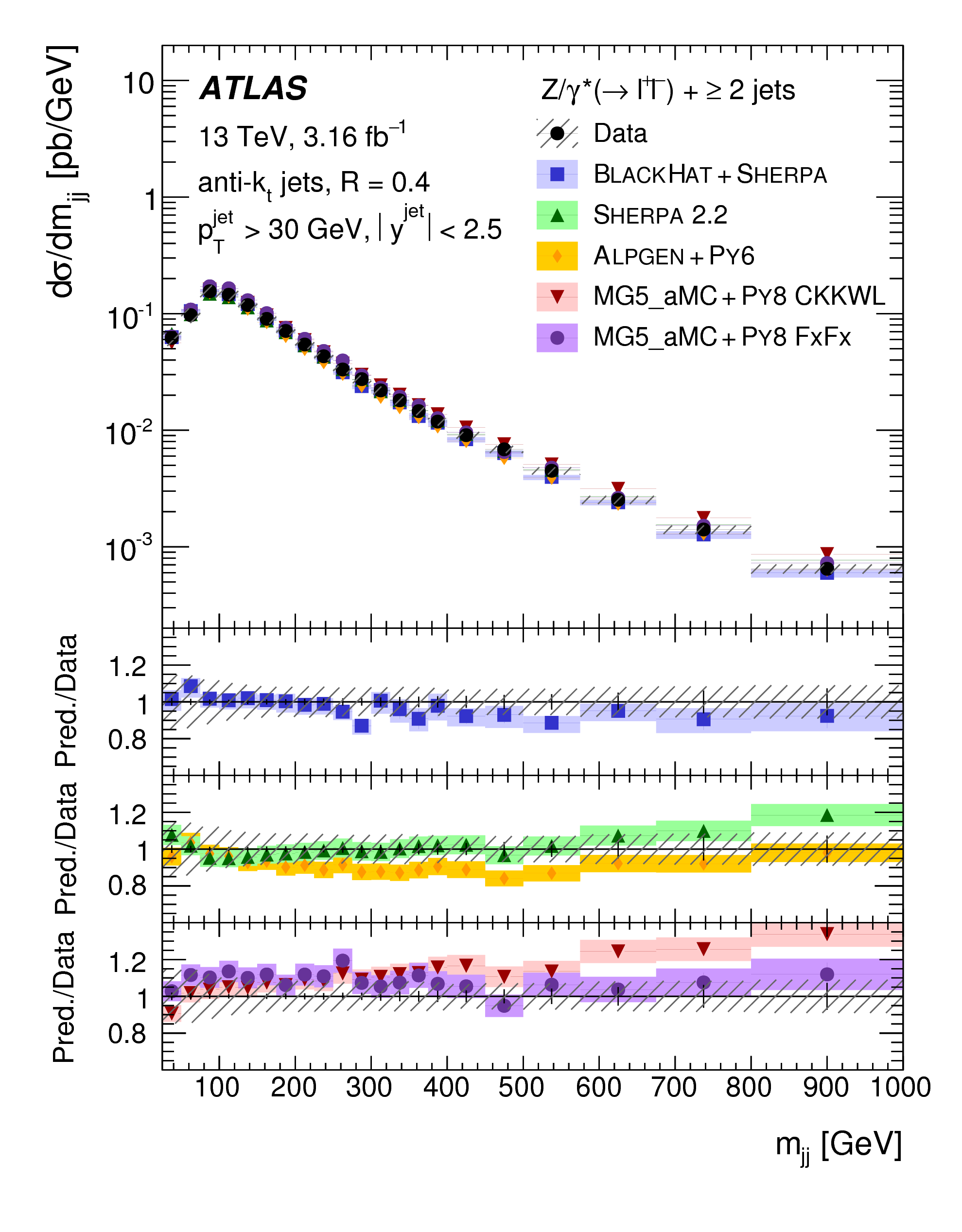} \\
\end{tabular}
}
\caption{On the left normalised differential cross section as a function of the invariant mass of the two leading jets in events of the electroweak enhanced region within the inclusive selection for electroweak $Wjj$ production described in Ref.~\cite{citeEWWjj}. On the right differential cross section as a function of the  di-jets invariant mass in a standard selection of a gauge boson ($Z$) and two jets described in Ref.~\cite{citeZjets}. \label{figMinv2pap}}
\end{figure}
Notice that in $Wjj$ EW enhanced region this variable is sensitive to CP violating anomalous trilinear gauge couplings. Hence good modelling of the SM physics is mandatory.  

The invariant mass of the di-jet system is shown on the left of Figure \ref{figMinv2pap} in the signal region of the VBF phase space probed by the $Wjj$ analysis;  on the right the corresponding cross section measured in the standard $Z$+jets analysis is displayed. A reasonable agreement between data and theory predictions is observed in the $Wjj$ case, provided the EW contributions are properly taken into account. In the inclusive $Zjj$ measurement, both NLO predictions and MC generators are in reasonable agreement with data, with the expected exception of the LO {\tt Madgraph} prediction, which overestimates data. However, the trend beyond 1~TeV deserves further attention as discussed in the next Section. 
  
\subsection{Electroweak production of $Zjj$ at 13 TeV }
\label{eqZjjprod}
Finally  some ATLAS results \cite{citeEWZjj13} on the EW production of $Zjj$ at 13 TeV are summarised here. Also in this case a general VBF-style selection is defined and the signal region is complemented by a QCD enhanced control region. 
\begin{figure}[thb]
\centerline{%
\begin{tabular}{cc}
\includegraphics[width=6.0cm]{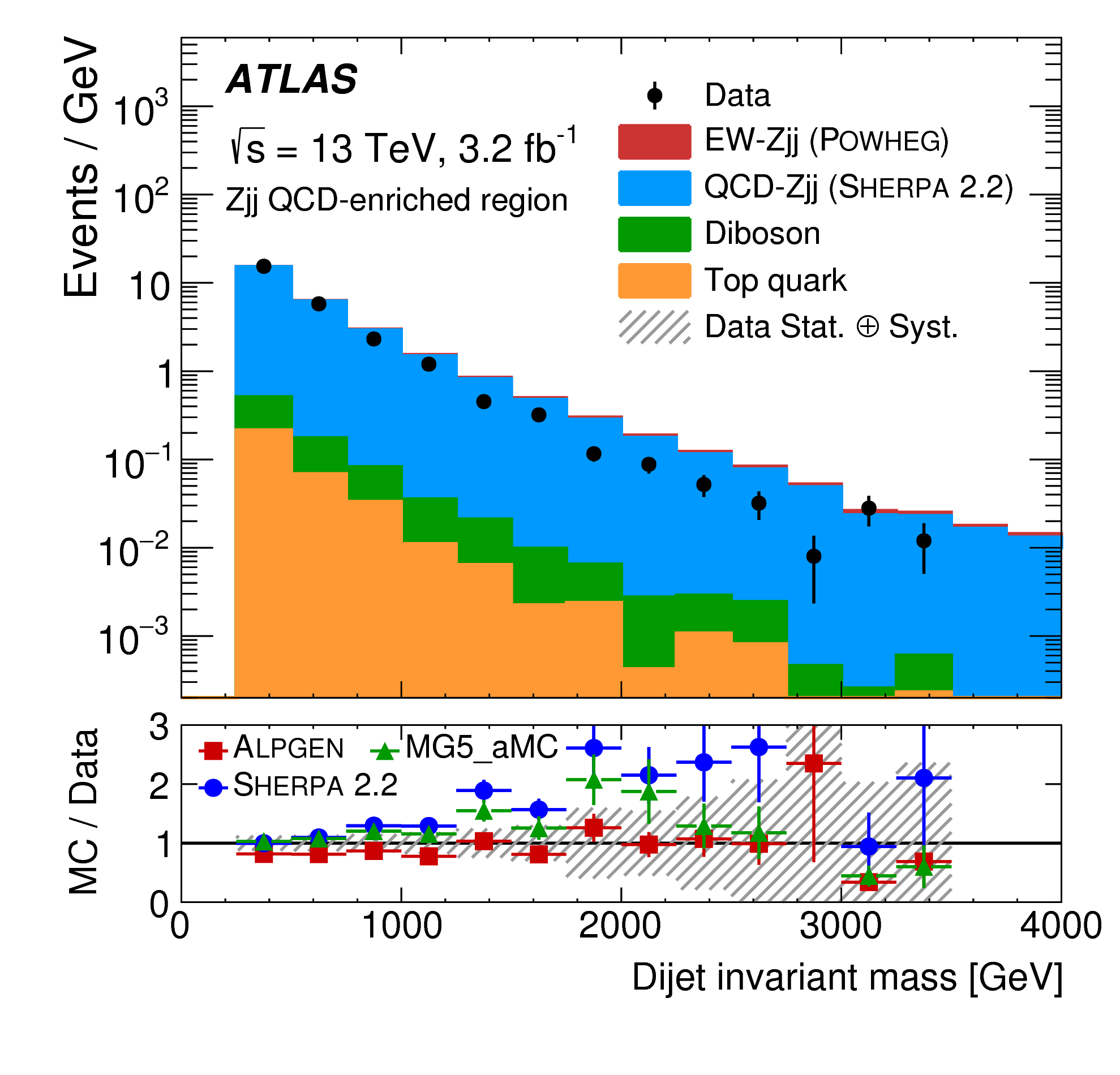} &  
\includegraphics[width=6.0cm]{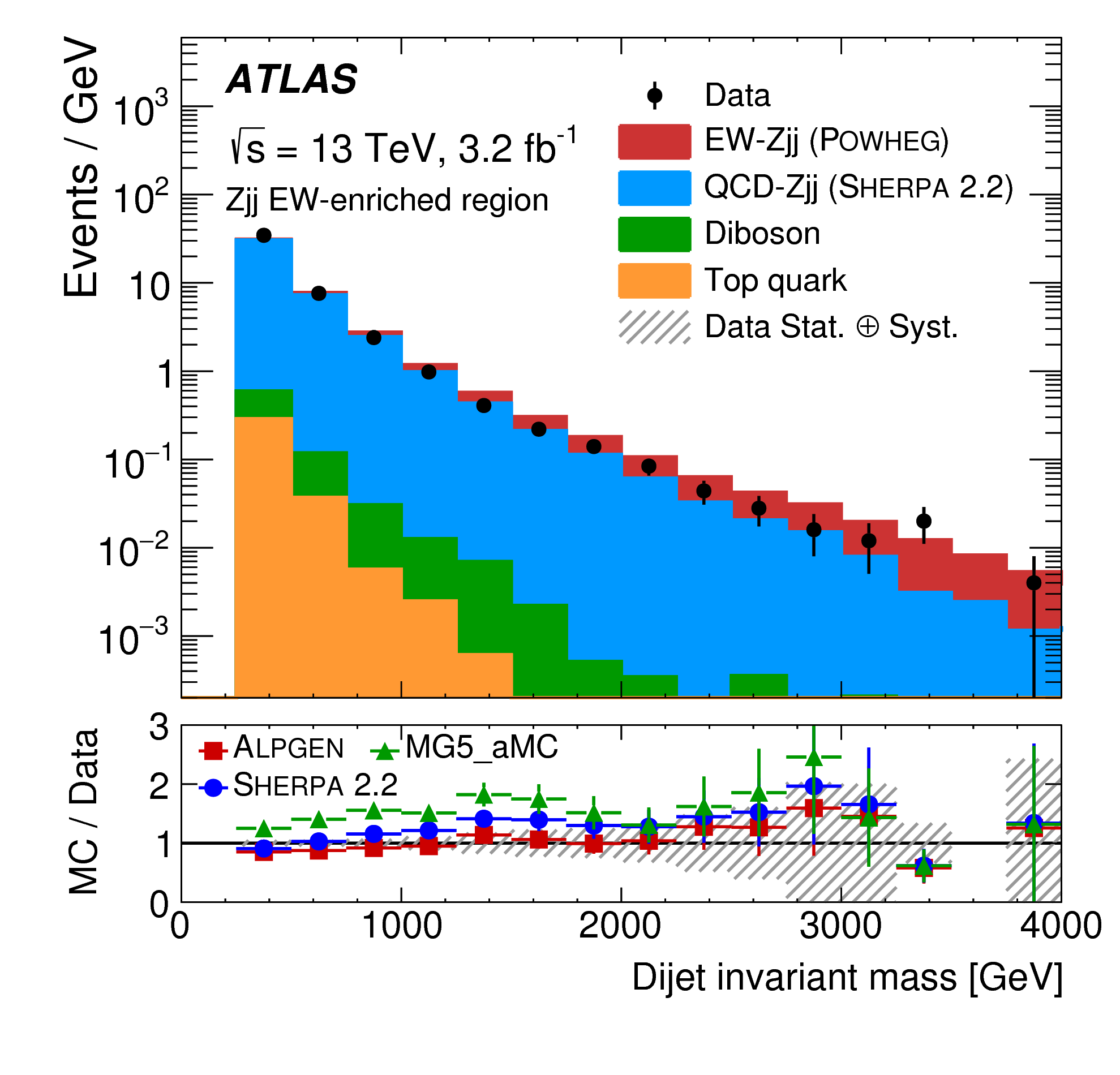} \\  
\includegraphics[width=6.0cm]{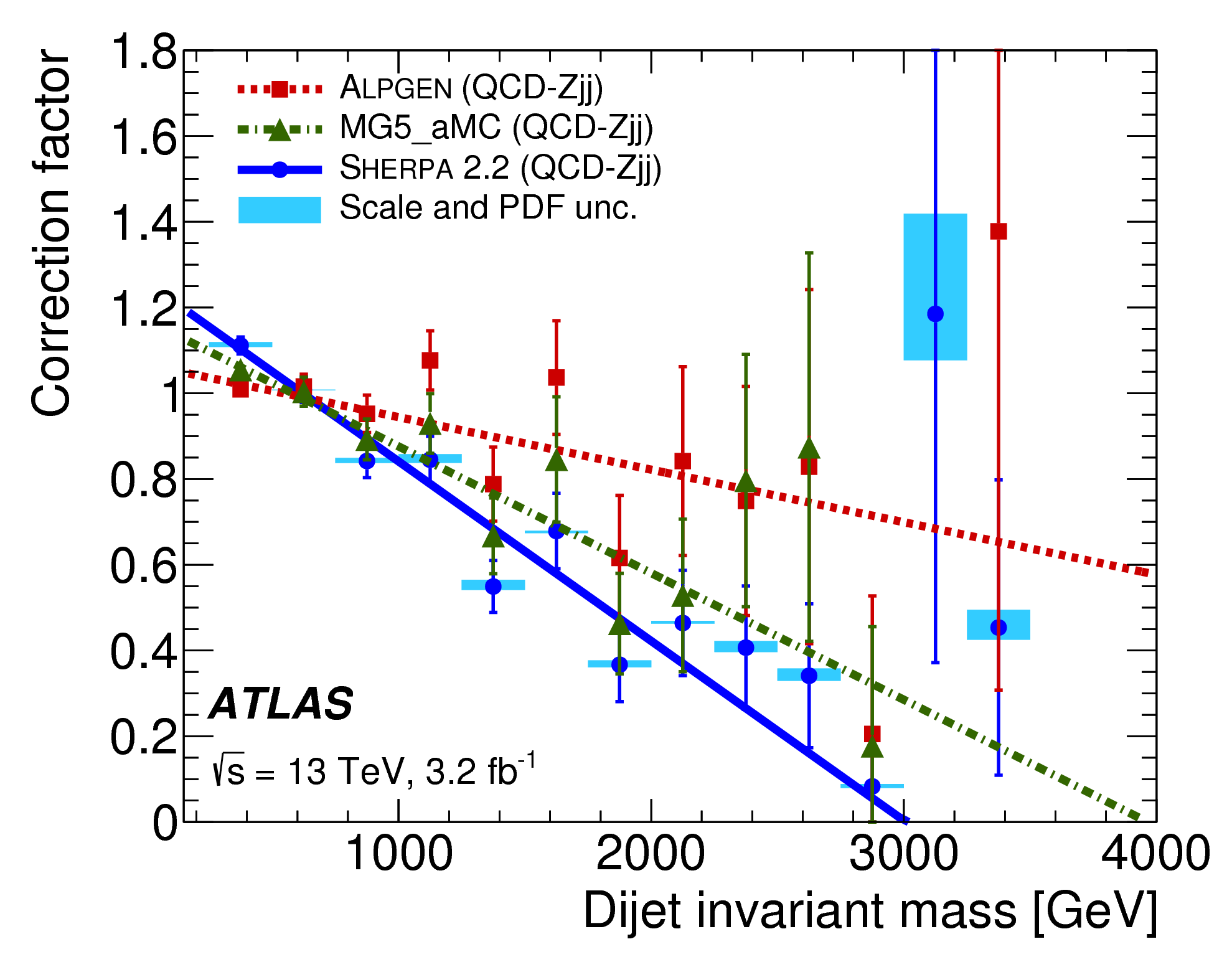} &  
\includegraphics[width=6.0cm]{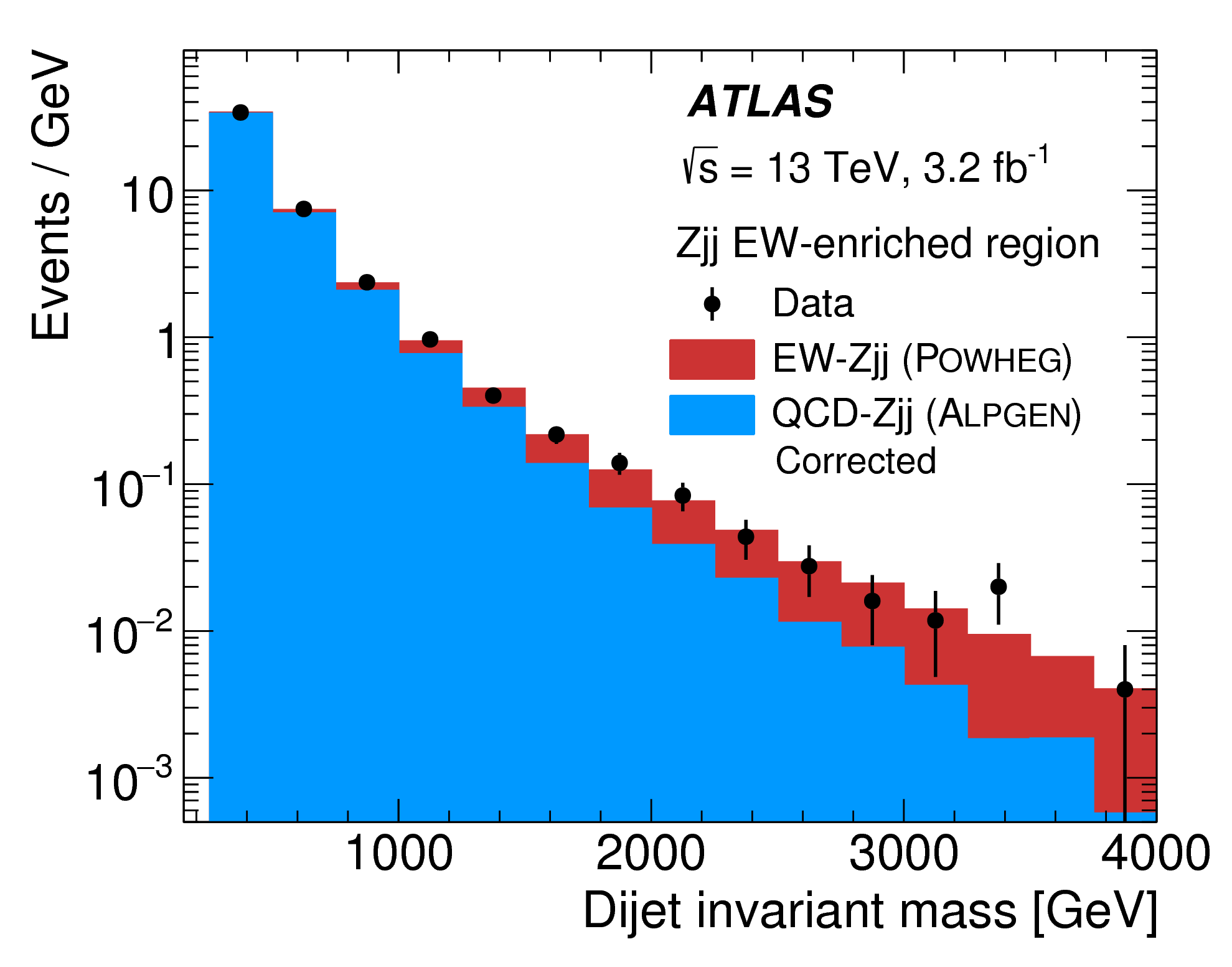} \\
\end{tabular}
}
\caption{Di-jet invariant mass distributions in data and MC in the selection for $Z$ + 2 jets produced via electroweak interactions presented in Ref.~\cite{citeEWZjj13}, for the QCD control region (top-left) and the signal region (top-right). In the bottom-left plot the correction factor between data and several MC generators derived from the QCD enhanced control region. In the bottom-right plot the di-jet invariant mass distribution for the signal region after applying the correction factor to {\tt Alpgen}.  \label{figMinv2ewZjj}}
\end{figure}
Figure \ref{figMinv2ewZjj} shows the di-jet invariant mass spectrum observed in the data for the QCD $Zjj$ CR, on the top left, and for the EW signal region, on the top right. Sherpa 2.2 is  failing to describe the spectrum of the dominant QCD component. Also other MC generators provide a poor modeling the di-jet invariant mass above 1 TeV; therefore, a data-driven correction, shown in the bottom left plot, is computed from the ratio of data and MC in the QCD enhanced CR. After applying the correction to the QCD prediction in the signal region, a nice agreement between data and MC is achieved, as demonstrated in the bottom right plot in Figure~\ref{figMinv2ewZjj}.
This need for theory improvements was suggested already in the ATLAS analysis of the electroweak production of $Z$ + 2 jets \cite{citeEWZjj8} based on the data at 8 TeV, published in 2014.

\section{Conclusions}
The selection of results summarised here gives the flavour of the rich program of SM measurements being performed by ATLAS in the boson plus jets domain. 
In many cases the data are well described by the QCD theory predictions, but that's not always the case.  In particular, a need for better understanding and modelling of the data is observed in the regions of the phase space that are accessed now at LHC for the first time (like high moment transfer in VBF topologies) and when properties of data are investigated from an unconventional perspective, like in the study of the \kt\ splitting scale. 
The precision of the available and foreseen measurements of gauge bosons plus jets at LHC has motivated a recent extensive study \cite{citeSimulVjets} of the MC generators describing these processes. Careful comparisons of the predictions  in various regions of the phase space, with and without focus on  electroweak Vjj production mechanisms and with or without additional requirements on the heavy-flavour content of the accompanying jets, show the relevance of higher-order corrections and of systematic theory uncertainties.


\bibliographystyle{apsrev4-1}


\end{document}